\documentclass[aps,prx,reprint,floatfix,superscriptaddress,footinbib,twocolumn]{revtex4}

\usepackage{xcolor}
\usepackage{microtype}
\definecolor{webblue}{rgb}{0, 0, 0.5} % less intense blue

\usepackage[bookmarks=true,
bookmarksnumbered=false, % true means bookmarks in
bookmarksopen=false, % true means only level 1
colorlinks=true,
linkcolor=blue]{hyperref}

\hypersetup{colorlinks=true,urlcolor=webblue,citecolor=blue}
\usepackage{graphicx}
\usepackage{subfigure}
\usepackage{url}
\usepackage{hyperref}
\usepackage{inconsolata}
\usepackage{amsmath}
\usepackage[capitalize]{cleveref}
\usepackage{braket}
\usepackage{outlines}
\usepackage{enumitem}
\usepackage{soul}
\usepackage{color}

\usepackage{bm} 

\usepackage[utf8]{inputenc}
\usepackage{siunitx,booktabs}
\usepackage{multirow}
\usepackage{amssymb}

\usepackage{relsize}

\usepackage{soul}
\usepackage{tabularx}

\usepackage{amsmath}
\usepackage{amssymb}
\usepackage{bbm}
\usepackage{braket}
\usepackage{xcolor}

\usepackage{comment}
\usepackage{pifont}

\begin{document}

\title{Real-Space Imaging of the Band Topology of Transition Metal Dichalcogenides}

\author{Madisen Holbrook}
\affiliation{Department of Physics, Columbia University, New York, NY, 10027, USA}

\author{Julian Ingham}
\affiliation{Department of Physics, Columbia University, New York, NY, 10027, USA}

\author{Daniel Kaplan}
\affiliation{Department of Physics and Astronomy, Center for Materials Theory, Rutgers University, Piscataway, New Jersey 08854, USA}

\author{Luke Holtzman}
\affiliation{Department of Applied Physics and Applied Mathematics, Columbia University, New York, New York 10027, USA}

\author{Brenna Bierman}
\affiliation{Department of Materials Science and Engineering, University of Wisconsin; Madison, USA}

\author{Nicholas Olson}
\affiliation{Department of Chemistry, Columbia University, New York, NY, USA}

\author{Luca Nashabeh}
\affiliation{Department of Physics, Columbia University, New York, NY, 10027, USA}

\author{Song Liu}
\affiliation{Department of Mechanical Engineering, Columbia University, New York, NY, 10027, USA}

\author{Xiaoyang Zhu}
\affiliation{Department of Chemistry, Columbia University, New York, NY, USA}

\author{Daniel Rhodes}
\affiliation{Department of Materials Science and Engineering, University of Wisconsin; Madison, USA}

\author{Katayun Barmak}
\affiliation{Department of Applied Physics and Applied Mathematics, Columbia University, New York, New York 10027, USA}

\author{James Hone}
\affiliation{Department of Mechanical Engineering, Columbia University, New York, NY, 10027, USA}

\author{Raquel Queiroz}
\affiliation{Department of Physics, Columbia University, New York, NY, 10027, USA}

\author{Abhay Pasupathy}
\affiliation{Department of Physics, Columbia University, New York, NY, 10027, USA}
\affiliation{Condensed Matter Physics and Materials Science Division, Brookhaven National Laboratory; Upton, NY, USA}

\date{\today}

\begin{abstract}
The topological properties of Bloch bands are intimately tied to the structure of their electronic wavefunctions within the unit cell of a crystal. Here, we show that scanning tunneling microscopy (STM) measurements on the prototypical transition metal dichalcogenide (TMD) semiconductor WSe$_2$ can be used to unambiguously fix the location of the Wannier center of the valence band. Using site-specific substitutional doping, we first determine the position of the atomic sites within STM images, establishing that the maximum electronic density of states at the $K$-point lies between the atoms. In contrast, the maximum density of states at the $\Gamma$ point is at the atomic sites. This signifies that WSe$_2$ is a topologically obstructed atomic insulator, which cannot be adiabatically transformed to the trivial atomic insulator limit.

\end{abstract}
 \maketitle

The wavefunctions of the electronic bands that govern the properties of crystalline solids are constructed from the building blocks of local atomic orbitals~\cite{Hoffmann1987HowChem,Blount1962formalisms}. The spatial structure of wavefunctions within a unit cell encodes many important properties -- for example, the electron density of an ionic solid resides on the atomic sites, distinguished from the covalent case where it is localized between them. In reciprocal space, the orbital character of a Bloch band can evolve across the Brillouin zone, along with the real-space structure of the wave functions. Recent work \cite{bradlyn2017topological,Cano2018building} established that the topological properties of a Bloch band can be classified by the transformation of the wavefunctions at high-symmetry points in the Brillouin zone with respect to the crystal symmetries. Since the spatial structure of the wavefunctions is directly related to their orbital content, this implies that probing the wavefunctions within the unit cell could reveal the topological properties of a solid. Although wavefunctions are not observables, the resulting electronic probability density can be directly probed within the unit cell with scanning tunneling microscopy (STM). However, relating the charge density to topological invariants such as the Berry phase \cite{Vanderbilt1993polarizationcharge} is challenging: the integrated charge density does not contain the vital phase information necessary to make this connection. For this reason, real-space probes such as STM are generally not considered effective tools for investigating the topological character of translationally invariant solids.

In this work, we show that precise STM measurements of the probability density at \textit{multiple energies} within a Bloch band can uncover the nature of the underlying wavefunctions. To make this connection, we examine Wannier functions (WF) of isolated electronic bands. WFs are a basis of local orbitals which parametrize a Bloch band, and accurately capture symmetries and local properties such as orbital hybridization and bonding. The WFs can be explicitly obtained via a Fourier transformation of the momentum-space Bloch states. WFs of covalent bands are centered at the bonds between atomic sites, distinct from ionic counterparts centered on the atomic sites. There is often no adiabatic transition between the two limits, and the WF is pinned to either the bond or the atomic site due to a symmetry-protected quantization of the Berry phase. The topologically distinct case in which WFs are centered at the bond positions is known as an ``obstructed atomic insulator" (OAI) \cite{bradlyn2017topological,Cano2018building,Khalaf2021boundary, Cano2021TQCrev,Xu2024filling}. Under certain circumstances, a careful measurement of the intra-unit cell charge density allows us to extract the structure of WFs and the associated Bloch wave function symmetries \cite{bauernfeind2021design, villanova2024ghost, nag2024pomeranchuk}. In this work, we utilize STM to probe the global topological properties hidden in the valence band of transition metal dichalcogenides (TMDs), wedding the delocalized picture of the band structure with the real-space view of the WF.

\begin{figure*}[t!]
    \centering
    \includegraphics[width=0.8\textwidth]{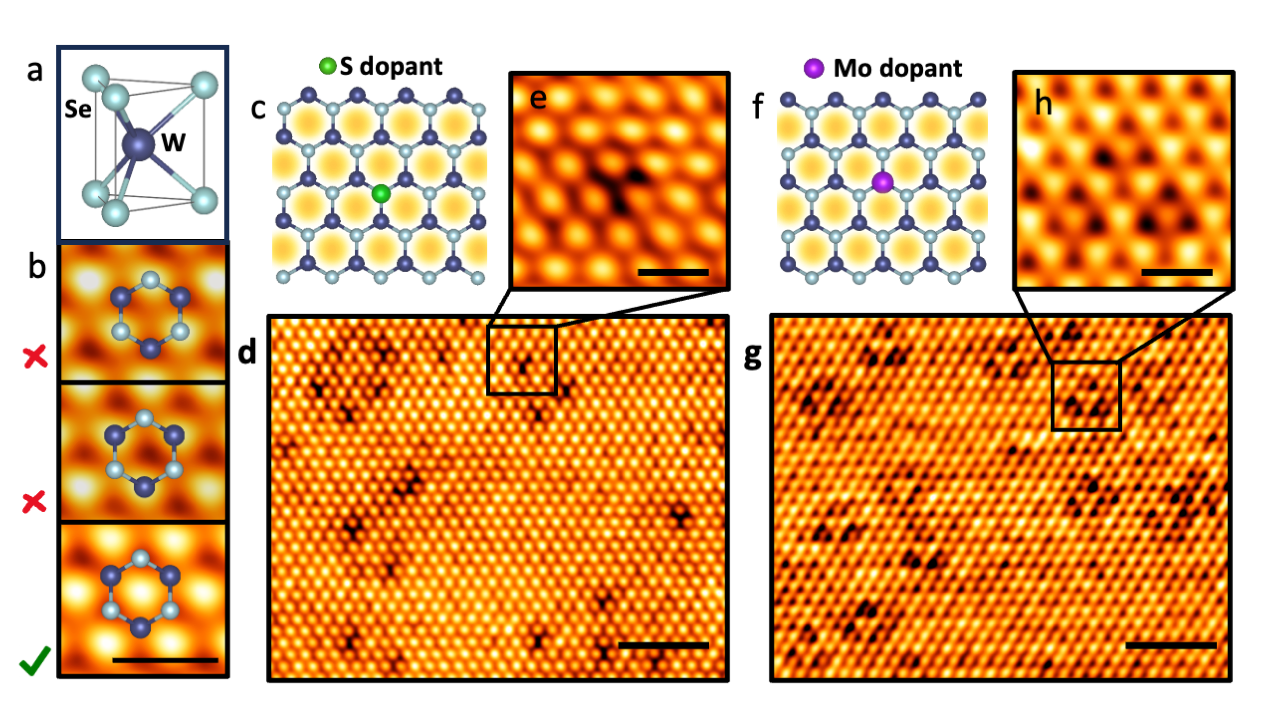}
    %[width=1\linewidth]
    \vspace{-0.2cm}
    \caption{\textbf{Connecting STM images to the WSe$_2$ atomic lattice.} (a) WSe$_2$ atomic structure (b) STM image of monolayer WSe$_2$ (c,f) Schematics showing a S, Mo dopant in WSe$_2$, respectively (d,g) STM images showing S and Mo doped WSe$_2$, respectively. (e,h) Zoom-in STM images of the S and Mo dopants, respectively. The scale bar in (b,e,h) is 0.5 nm, and (d,g) is 2 nm. Sample bias in V is (b) -1.4, (e) -1.7 and (d, g, h) -1.8.}
    \label{f:fig1}
    \vspace{-0.1cm}
\end{figure*}

TMDs of the H polytype have a trigonal prismatic configuration of transition metals sandwiched between the top and bottom layers of chalcogen atoms, Fig. \ref{f:fig1}a. Our STM investigation centers on mechanically exfoliated WSe\textsubscript{2} monolayers obtained from high-purity single crystals \cite{Liutwostep} on graphite substrates prepared by conventional dry transfer techniques \cite{Li1dcontact} (see Methods). An STM image of monolayer WSe\textsubscript{2} (Fig. \ref{f:fig1}b) reveals a periodic structure resembling a hexagonal atomic lattice. However, identifying the atomic positions is nontrivial because the metal, chalcogen, and hollow lattice sites all share the same hexagonal symmetry (Fig. \ref{f:fig1}b), which led to debate in early STM studies of TMDs \cite{Stupian1987Matoms,Weimar1988Xatoms, WHANGBO1995}. Localized charge features observed by STM are often associated with the atoms that predominantly contribute to the band edges, as exemplified in GaAs \cite{Feenstra1987GaAs}. However, in the 2H TMDs, two opposing considerations come into play. On the one hand, the transition metal $d$ orbitals dominate the band edges. On the other hand, the exponential sensitivity of tunneling often implies that the closest atoms to the tip are observed by STM. Therefore, several previous theoretical studies attributed the features in STM images to the surface chalcogen atoms \cite{ALTIBELLI1996, KOBAYASHI1996, WHANGBO1995}. However, comparison of STM and non-contact atomic force microscopy images of defects in TMDs hints that the bright contrast features in STM images might not correspond to chalcogen sites \cite{barja2019identifying}.

In order to conclusively identify the lattice positions in the STM images, we introduce substitutional dopants at the metal and chalcogen sites. Importantly, the high purity of the self-flux grown WSe\textsubscript{2} makes it straightforward to differentiate the dopants from the sparse native defects \cite{Liutwostep} (see Methods). As shown schematically in Fig. \ref{f:fig1}c, we first replace the Se atoms with isovalent substitutional S dopants to pinpoint the Se lattice sites. Fig. \ref{f:fig1}d shows a valence band STM image of the S doped WSe\textsubscript{2}, revealing a high density of defects. These defects are centered \textit{between} the bright features in the topograph (see Fig. \ref{f:fig1}e zoom-in). Thus, contrary to the commonly held belief, the ``atoms" seen in STM images of WSe\textsubscript{2} do not correspond to the chalcogen sites.

Mirroring the S-doping approach, we next determine the W lattice sites by substitutionally doping with Mo (Fig. \ref{f:fig1}f). Fig. \ref{f:fig1}g demonstrates that the Mo defects are also centered between the bright topographic maxima of the STM image (see Fig. \ref{f:fig1}h zoom-in). Therefore we conclude that the bright spots seen in the STM images at this bias voltage do not correspond to either the S or the W atoms in WSe\textsubscript{2}, but instead are located in the center of the honeycomb or ``hollow" sites.

\begin{figure*}[t!]
    \centering
    \includegraphics[width=0.77\textwidth]{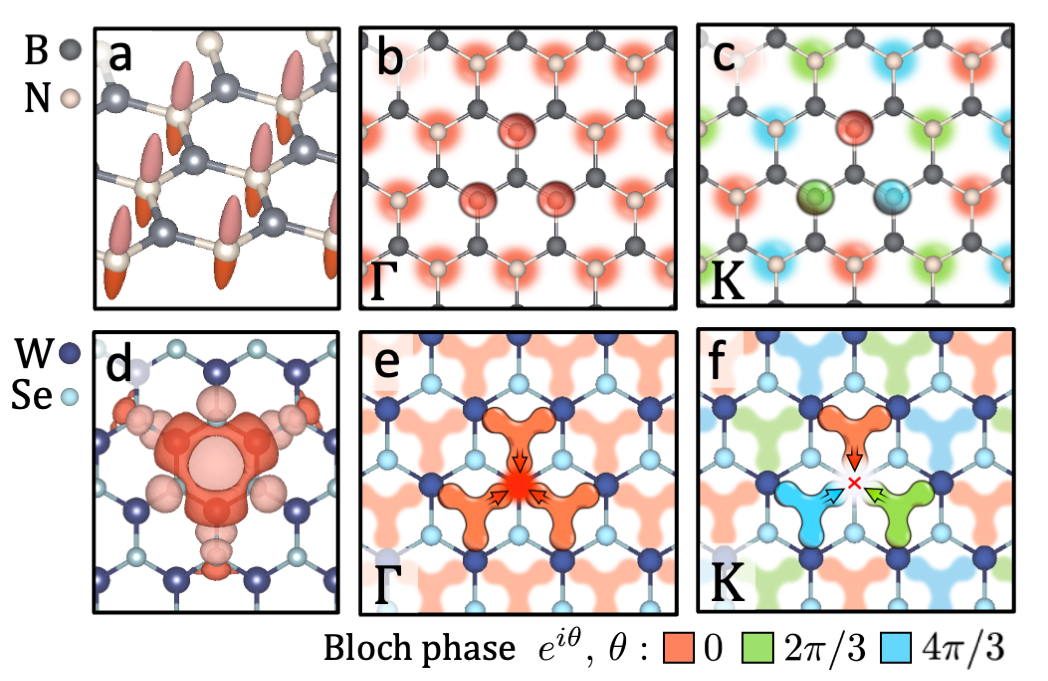}
    %[width=1\linewidth]
    \vspace{-0.1cm}
    \caption{\textbf{Charge density profile from interference of Wannier orbitals in 2H TMDs.} (a) DFT calculation of the maximally localized Wannier function of the hBN valence band centered on the N atoms. (b,c) Schematic showing the hBN Bloch states at $\Gamma$ and $K$, respectively. The Bloch phase modulation is represented by red, green, and blue colors (see key). (d) Maximally localized Wannier function of the WSe\textsubscript{2} valence band centered between the W atoms. (e,f) Schematic showing the WSe\textsubscript{2} Bloch states at $\Gamma$ and $K$ showing constructive and destructive interference of the Bloch states at the W sites, respectively.} 
    \label{f:fig2}
    \vspace{-0.1cm}
\end{figure*}

To understand this observation, we analyze how Bloch states are constructed from the building blocks of WFs in two distinct limits: the trivial case of ionic hexagonal boron nitride (hBN), and an OAI exemplified by WSe\textsubscript{2}. Both crystals have band gaps with a single relevant valence band that has one occupied electron per spin. Viewed from a two dimensional plane, both crystals similarly have unit cell vectors $\mathbf{a}_\pm=a/2(\sqrt{3}\mathbf{x}\pm\mathbf{y})$, $C_3$ rotation symmetry, and mirror plane symmetry, with three distinct maximal Wyckoff positions (WP). These three WP correspond to the locations of the W (N) atoms, Se (B) atoms, and hollow sites, respectively, and each appears once in the unit cell with $C_3$ rotation symmetry. This symmetry requires that for both cases, the WF occupies one of the distinct WPs and transforms trivially with no phase change under $C_3$ rotation. The important difference in the bonding in these crystals lies in where -- \textit{i.e.} at which WP -- the WF is centered.

We first examine the hBN valence band, which is predominantly composed of the N-$p_z$ orbitals. We construct a maximally localized Wannier function spanning the majority of the valence band (Suppl. Mat. Fig. S3), shown in Fig. \ref{f:fig2}a (see Methods). The WFs closely resemble the constituent $p_z$ orbitals and transform trivially under $C_3$. Notably, they are extremely localized to the N sites with a significantly smaller spread $\sigma = 0.43a$ than the unit cell, Fig. \ref{f:fig2}a. The Bloch states at each momentum $\mathbf{k}$ can be described by a superposition of Wannier orbitals decorated with momentum-dependent Bloch phases, 
\begin{align}
    \psi_{\mathbf{k}}(\mathbf{r}) = \sum_{\mathbf{\mathbf{R}}} e^{i\mathbf{k}\cdot\mathbf{R}} w(\mathbf{r-R}),
\end{align}
where $\mathbf{R}$ are the WF centers. The Bloch phase modulation of the WF is shown schematically in Fig. \ref{f:fig2}c and d, represented by the red, green, and blue colors. There is no phase difference at the $\Gamma$-point (Fig. \ref{f:fig2}c), while at the $K$-point (Fig. \ref{f:fig2}d) the Bloch phase $e^{i\mathbf{K}\cdot\mathbf{r}}$ differs by $e^{\pm2\pi i/3}$ for sites separated by the unit cell vectors $\mathbf{a}_\pm$, and both states transform trivially under $C_3$ symmetry with respect to the N site. Importantly, regardless of Bloch phase, the charge density is localized at the N sites due to the lack of overlap of the tightly localized WF. Therefore, a hypothetical STM measurement would show a peaked density of states at the N sites at all energies probed within the valence band. 

We now consider the contrasting example of WSe\textsubscript{2}. In this case the valence band is primarily composed of strongly hybridized $d_{x^2-y^2}$, $d_{xy}$ and $d_{z^2}$ orbitals of the W atoms. At the $\Gamma$ point, the band is dominated by $d_{z^2}$ character that has a trivial eigenvalue under $C_3$ transformation. In contrast, the $d_{x^2-y^2}\pm id_{xy}$ character of the $K$/$K'$ points transforms under a chiral representation with $C_3$ eigenvalue $e^{\pm2\pi i/3}$. Using the band representations of this symmetry group, we can infer that this combination of $C_3$ eigenvalues is incompatible with orbitals localized at the W sites, whose $C_3$ representation is momentum-independent \cite{bradlyn2017topological}. DFT calculation of the maximally localized WF (see Methods) shown in Fig. \ref{f:fig2}d, supports this conclusion, showing that the WF is indeed centered at the hollow site between the W atoms, with a spread $\sigma = 2.89 a$ that exceeds the unit cell. In Fig. \ref{f:fig2}e and \ref{f:fig2}f we represent Bloch states similar to the hBN example. At $\Gamma$, the Bloch state transforms trivially with $2\pi/3$ rotation about the W site, while at $K$, there is an added phase of $e^{2\pi i/3}$ indicated by the colors in Fig. \ref{f:fig2}f. The lack of Bloch phase modulation at the $\Gamma$-point leads to constructive interference at the W site where the Wannier orbitals overlap, Fig. \ref{f:fig2}e. In contrast, the Wannier orbitals add destructively at the $K$-point, and a node in the charge density appears at the W site, Fig. \ref{f:fig2}f. This destructive interference at the W site is therefore a defining signal of the nontrivial $C_3$ eigenvalue at $K$.

\begin{figure*}[t!]
    \centering
    \includegraphics[width=0.7\textwidth]{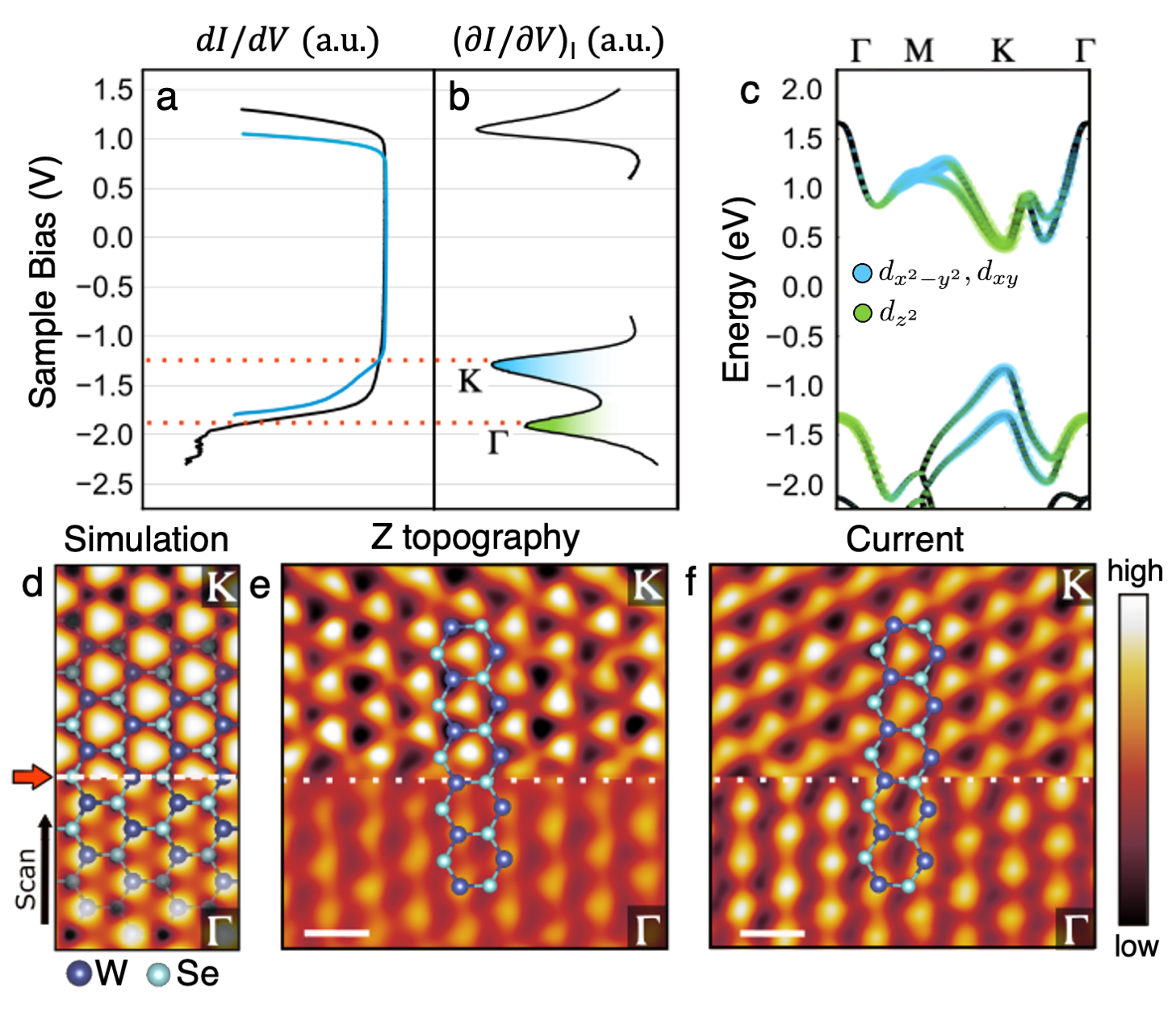}
    %[width=1\linewidth]
    \vspace{-0.4 cm}
    \caption{\textbf{Bias-dependent imaging of WSe$_2$.} (a) Constant height $dI/dV$ spectroscopy obtained with ranges that contain (black) and exclude (blue) $\Gamma$ tunneling (b) Variable z $(\partial{I}/\partial{V})\textsubscript{I}$ spectroscopy with the $K$ and $\Gamma$-points indicated by red dashed lines. (c) DFT calculation of the WSe\textsubscript{2} band structure with the orbital weight of $d_{x^2-y^2}$/$d_{xy}$ (blue), and $d_{z^2}$ (green). (d) First principles (partial charge density) STM simulation of the charge density at the VBM (top) and past $\Gamma$ (bottom), showing a shift in the charge density position with the bias change (red arrow, white line) (e,f) Constant current Z topography and constant height current STM image of WSe\textsubscript{2} with a bias change (white dashed line) with the atomic lattice overlaid showing that the high topography/current positions shifts from the W site (bottom) to the honeycomb center (top) when the bias is changed. Sample bias in V in (e) is -2 V (bottom) to -1.4 V (top) and (f) -1.9 V (bottom) to -1.4 V (top). The scale bars in (e,f) are 0.4 nm}
    \label{f:fig3}
    \vspace{-0.1cm}
\end{figure*}

This simple picture not only explains our above STM observations (Fig. \ref{f:fig1}), but also predicts that measurements deeper in the band toward the $\Gamma$ point might show the charge density shift from the hollow site toward the W atoms. In this space group, identifying the difference in the symmetry eigenvalues at $\Gamma$ and $K$ high symmetry points is sufficient to diagnose the Wyckoff position where the Wannier functions are centered, as is known from the theory of symmetry-based indicators \cite{bradlyn2017topological}. Therefore, observing this charge density shift due to the different symmetry of the $\Gamma$ and $K$ Bloch states provides direct evidence that WSe\textsubscript{2} is an obstructed atomic insulator with a quantized Berry phase. 

 To explore this phenomenon, we first determine the energy locations of the high symmetry points in the WSe\textsubscript{2} valence band using $dI/dV$ spectroscopy. Probing the VBM at the $K$-point in monolayer TMDs is only possible at smaller tip-sample distance due to the increase of the tunneling decay constant with larger in-plane momentum \cite{ZhangCritical}. Fig. \ref{f:fig3}a shows conventional $dI/dV$ spectroscopy obtained with a constant tip-sample distance at a smaller bias range where the tip is closer and sensitive to $K$ tunneling (blue curve). In contrast, the $K$ states are not resolved in larger bias ranges where the tunneling current is dominated by the $\Gamma$ point (black curve). We follow a previous approach to address this issue by implementing constant current $dI/dV$ spectroscopy, which allows the tip height to vary across the different tunneling regimes \cite{stroscio1993}, Fig.~\ref{f:fig3}b. The peaks seen in this spectrum correspond to peaks in the density of states at the high symmetry points, and we can assign the WSe\textsubscript{2} VBM at $K$ to -1.24 eV and $\Gamma$ to -1.89 eV, shown in Fig. \ref{f:fig3}b with band structure aligned for comparison in Fig. \ref{f:fig3}c.

\begin{figure*}[t!]
    \centering
    \includegraphics[width=0.8\textwidth]{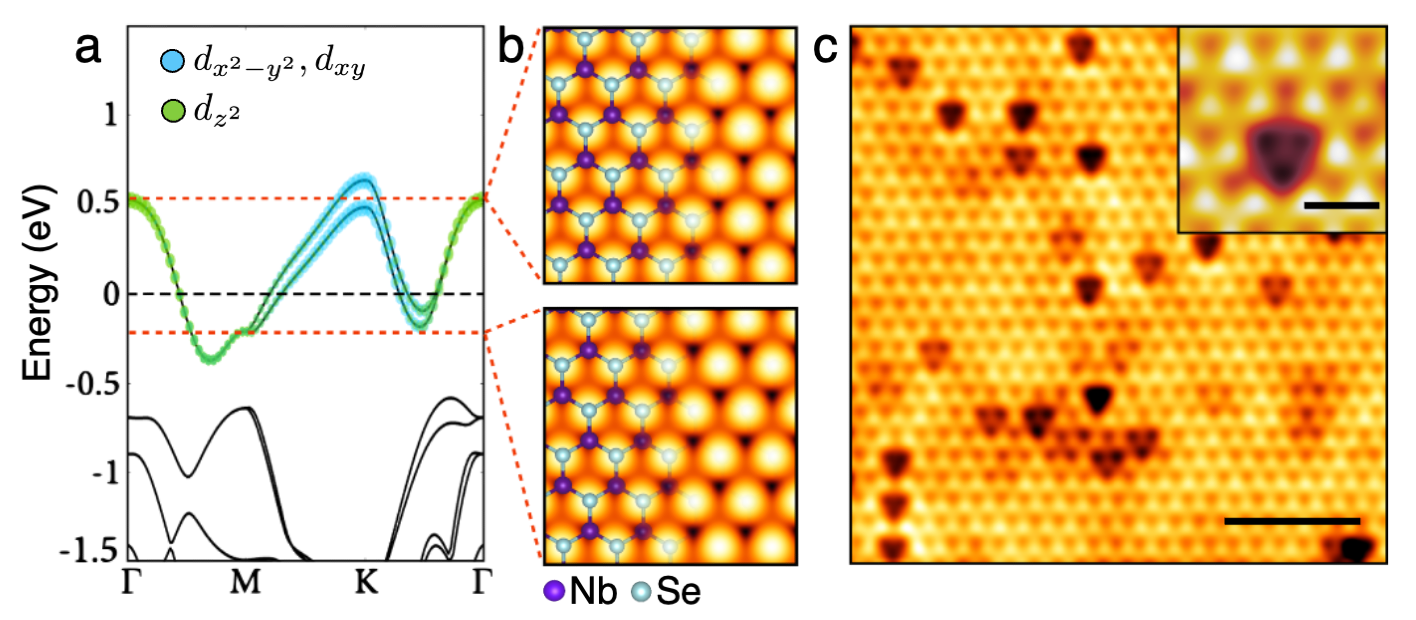}
    %[width=1\linewidth]
    \vspace{-0.2cm}
    \caption{\textbf{Atomically centered charge density in NbSe$_2$.} (a) First-principles calculations of NbSe$_2$ band structure showing the orbital weight of $d_{x^2-y^2}$/$d_{xy}$ (blue), and $d_{z^2}$ (green) of NbSe$_2$. This highlights the same orbital composition as in WSe$_2$, but the $\Gamma$ and $K$ points have nearly degenerate energies. (b) Simulated STM images show the bias-dependence of the charge density. The overlaid atomic lattice shows that the charge density is localized on the Se sites at the positive and negative energies, indicated by red dashed lines in (a). (c) STM image of S doped NbSe$_2$ shows the S defects (inset) are centered on the bright contrast features, indicating they correspond to the Se sites of the NbSe$_2$ lattice. The bias voltage in (c) is 0.89 V and the scale bars are 2 nm and 0.5 nm in the inset.}
    \label{f:fig4}
    \vspace{-0.1cm}
\end{figure*}

With the energy positions of the WSe\textsubscript{2} valence band established, we conceive a simple procedure to probe the energy dependence of the charge density using STM, shown schematically in (Fig. \ref{f:fig3}d). We begin a scan with a sample bias voltage below the $\Gamma$ point, then change to a voltage near the K point within the same scan frame. This strategy enables a direct comparison of the spatial features of the charge density within the atomic lattice at the same location without lateral drift, c.f. Supplementary Material (SM). A constant-current z topography image obtained in this way (Fig. \ref{f:fig3}e) reveals a clear shift in the corrugation after the bias voltage change (white dotted line). To remove this uncertainty of varying tip height, we obtain current images at different bias voltages with the feedback off, maintaining a constant tip height (Fig. \ref{f:fig3}f). The magnitude of the tunneling current changes significantly between the two bias regimes, therefore, we move the tip closer when probing $K$ to maintain the same current (see SM). Analysis of the current image reveals close agreement with the topography image, showing a consistent shift in the charge density. We determine the WSe\textsubscript{2} lattice alignment (overlay in Fig. \ref{f:fig3}e,f) using nearby defects (see SM), revealing that the charge density indeed moves from the W atoms to the honeycomb center when traversing the valence band from $\Gamma$ to $K$. Our STM investigations confirm that the location of the peak charge density evolves along with the orbital symmetry across the Brillouin zone, providing evidence that the WSe\textsubscript{2} valence band is obstructed.

Observing a discernible shift in the charge density might seem surprising, as all states from the band edge up to the bias setpoint contribute to the tunneling current in STM measurements. Many states contribute to this sum at $\Gamma$, including the states at the $K$ point. The ability to resolve the shift in the charge density stems from the unique characteristics of the WSe\textsubscript{2} valence band. First, the large energy separation between the $K$ and $\Gamma$ points allows for selective imaging of the VBM at $K$. Second, as seen in the $dI/dV$ spectroscopy (Fig. \ref{f:fig3}a blue, black), the tunneling current increases dramatically at $\Gamma$ and dominates the $K$ states due to the difference in their tunneling decay constants. Thus, we see that the WSe\textsubscript{2} valence band provides a convenient model system to separately probe the orbital symmetry of the high symmetry points. 

The obstruction in WSe\textsubscript{2} is not unique, and is predicted to be a general property within many 2H TMDs \cite{Liu2013three-band, Zeng2021multiorbital, Jung2022hiddenkagome}. This topological character ultimately results from the different symmetry eigenvalues at the high symmetry points of a band, and so it is impossible to probe the shift in the charge density if the energies at the high symmetry points are not well separated. Metallic NbSe\textsubscript{2} is a natural example of this subtlety -- the valence band possesses the same topological properties as WSe\textsubscript{2}, as we illustrate with first-principles calculations in Fig. \ref{f:fig4}a, but the Fermi level intersects both the $\Gamma$ and $K$ points. STM should, therefore, probe both pockets regardless of the energy, exhibiting a bright contribution to the density of states at the atomic sites, as shown by our STM simulations in Fig. \ref{f:fig4}b. Adopting the same substitutional doping strategy, we introduce S dopants to NbSe\textsubscript{2} to identify the Se sites. Figure \ref{f:fig4}b shows an STM image of the doped NbSe\textsubscript{2}, revealing a marked contrast to the WSe\textsubscript{2} case, with S defects centered at the bright positions. A zoom-in of one of the S defects (Fig. \ref{f:fig4}c) shows an obvious depression at the Se lattice site, unambiguously confirming that the STM images reflect surface Se atoms of NbSe\textsubscript{2}. Although the overlapping energies of the high symmetry points and metallic nature of NbSe\textsubscript{2} obscures its topological character, the Berry phase might be uncovered by measuring bias-dependent interference near translational symmetry-breaking defects \cite{Stern.Queiroz.2024,teo2010topological, ran2009one, slager2015impurity, jurivcic2012universal}, as has been explored in graphene \cite{dutreix2019measuring}.

Our STM observations of the obstructed topology of the WSe$_2$ can be equivalently understood as direct probing covalent bonds between metal sites that are occupied by valence electrons. This is at odds with the naive prediction that the electrons would be shared in covalent bonds between the metal and chalcogen sites, due to their similar electronegativity. The topological obstruction of WSe$_2$ has a number of physical ramifications -- including the existence of corner states \cite{Zeng2021multiorbital}, and an enhancement of the dielectric constant \cite{komissarov2024quantum}. In addition to elucidating the topology of 2H TMDs, our results clarify previous difficulties in identification of defects in these materials -- which have led to contradictory defect assignments, \cite{zhang2019engineering, barja2019identifying, lin2018realizing, trainer2022visualization} -- inviting a deeper examination of those previous results. 

In the context of finite-sized systems such as molecules \cite{Lee1999singlebond, Groos2011orbitimaging, Weiss1993Benzene} and quantum simulators \cite{Sierda2023, gardenier2020}, it is widely appreciated that the spatial profile of the density of states reflects the structure of molecular orbitals, which may be delocalized from the constituent atoms. By contrast, the potential disparity between atomic positions and peaks in the density of states is less well appreciated in the context of translationally invariant solids. Rather, STM images reflect the probability density of electronic orbitals combined with phases, which may interfere with one another depending on the probed momenta within a Bloch band. Charge density that is delocalized from the atomic lattice has been observed in other systems such as surface dimer bonding of Si \cite{Tromp1986Si001}, surface states of kagome metals \cite{villanova2024ghost, nag2024pomeranchuk}, moiré heterostructures \cite{zhang2024direct, thompson2024visualizing} and spin-Hall insulator indenene \cite{bauernfeind2021design, eck2022real}. Our study is the first to observe the variation of the charge density between different high symmetry points within the \textit{same} band of a periodic solid, and use this to diagnose its band topology. This contrasts with prior real space diagnostics of topology, which invoke signatures arising from translational symmetry breaking near defects, step edges, or sample boundaries \cite{nayak2019resolving, ugeda2018observation, yin2021probing, jeon2024resolving}. Our methodology introduces new avenues for probing band topology, providing insights into how real space observables can serve as indicators of underlying topological properties in quantum materials.

\section*{Acknowledgement}

This work was primarily supported by the NSF MRSEC program at Columbia through the Center for Precision-Assembled Quantum Materials (DMR-2011738). 
JI, DK and RQ acknowledge support by grant NSF PHY-2309135 to the Kavli Institute for Theoretical Physics (KITP). ANP acknowledges support by the Air Force Office of Scientific Research via award FA9550-21-1-0378. RQ and JI are further supported by NSF Career Award No. DMR-2340394. DK is supported by the Abrahams postdoctoral fellowship of the Center for Materials Theory, Rutgers University, and the Zuckerman STEM fellowship. 

\section*{Methods}

\textbf{TMD Synthesis.}  The pristine and 1\% Mo doped  WSe\textsubscript{2} crystals were synthesized using a self-flux method with a metal to chalcogen molar ratio of 1:100 following the method of Ref. W and Mo powder of 99.999\% and 99.997\% purity and Se shot of 99.999+\% purity were put in a quartz ampule that was evacuated and sealed at ~10-5 Torr. The ampules were heated to 1000 °C over 24 h, held at that temperature for 2 weeks, cooled at a rate of 1 °C/h to 500 °C, then cooled to room temperature at 5 °C/h. The  WSe\textsubscript{2} crystals and excess selenium were transferred to a new quartz ampule with quartz wool, evacuated, sealed, and heated to 285 °C to melt and separate the excess selenium by centrifuge. 

Single crystals of WSe$_{2-x}$S$_x$ were grown through chemical vapor transport using I$_2$ as the transport agent. W powder (99.999, Fisher Sci.), Se shot (99.999, Fisher Sci.), S pieces (99.9995, Fisher Sci.), and I$_2$ lump (99.999, Alfa) were loaded into a quartz tube with a ratio of 1:2:0.02 (W:Se:S), then sealed under vacuum ($\sim$1 mTorr). Subsequently, the ampule was placed in a two-zone furnace and heated to 1000 $^\circ$C over 12 h and a temperature gradient introduced over 2 h by increasing the temperature of one side of the furnace to 1150°C. The sample was held at this gradient for one week, then cooled (5.4 $^\circ$C/hr) to 550 $^\circ$C, before removing to room temperature. Harvested crystals were washed with isopropyl alcohol and acetone to remove the transport agent. 

The NbSe$_{2-x}$S$_x$ crystals were grown by the self-transport method starting from stoichiometric proportions of elemental Nb (99.997\%), Se (99.999+\%), and S (99.999+\%) powders over the period of one week in a temperature gradient of 950 to 900°C.\\

\textbf{Fabrication of Stacked Monolayer TMD Samples.} Bulk TMD and highly oriented pyrolytic graphite (HOPG) crystals were mechanically exfoliated using the conventional tape technique onto 285 nm SiO2/Si substrates. The monolayer WSe2/graphite heterostructures were assembled by the layer-by-layer dry stacking method with poly(propylene carbonate) (PPC) \cite{Li1dcontact}. The stacks were transferred to a SiO2 wafer by melting the polymer at 100 °C with the vdW stack on top. Indium contacts were then made to the graphite before the wafer was mounted on the STM sample holder using conductive silver paste (EPO-TEK H20E). \\

\textbf{STM Characterization.} STM measurements were obtained using a commercial Omicron ultrahigh vacuum STM system. A chemically etched W STM tip was cleaned and calibrated using an Au(111) single crystal prior to the measurements. Measurements were obtained at room temperature (300 K). \\

\textbf{DFT calculations}
The Vienna Ab-initio Simulation package (VASP) \cite{VASP1,VASP2} was used for obtaining the ground state charge density of WSe\textsubscript{2}, NbSe\textsubscript{2} and hBN. Lattice constants of $a = 3.32,3.45, 2.51$\AA~were used in each case, respectively. The exchange-correlation functional used was PBEsol \cite{PBEsol}. A plane-wave cutoff $E = 380 \textrm{eV}$ was applied. In the case of WSe\textsubscript{2}, NbSe\textsubscript{2}, the metal-chalcogen bonds were relaxed until forces were less than $1 \textrm{meV}$\AA$^{-1}$. The ground state charge density was then projected onto Wannier functions using Wannier90 \cite{wannier90}. 

\bibliography{references.bib}

\widetext

\clearpage

\begin{center}
\textbf{\large Supplementary Material}
\end{center}
\setcounter{equation}{0}
\setcounter{table}{0}
\setcounter{section}{0}
\setcounter{figure}{0}
%\makeatletter
\renewcommand{\theequation}{S\arabic{equation}}
\renewcommand{\thefigure}{S\arabic{figure}}
\renewcommand{\thesection}{S\arabic{section}}
%\renewcommand{\citenumfont}[1]{S#1}
%\begin{widetext}

\section{Model calculations of the local density of states (LDOS)}

The differential tunneling current imaged in STM is proportional to LDOS$(\text{eV},\mathbf{r})$, where eV is the bias voltage, and the local density of states (LDOS) is given by
\begin{align}
    \text{LDOS}(E,\mathbf{r}) = - \tfrac{1}{\pi} \text{Im} \langle c_\mathbf{r}c^\dag_\mathbf{r}\rangle_E = -\tfrac{1}{\pi} \text{Im} G(E, \mathbf{r})
\end{align}
where $c^\dag_\mathbf{r}$ is the electron creation operator and $G(E,\mathbf{r})$ is the electron Green' s function.  To relate the continuum electronic density of states to the wavefunctions of a lattice model, we expand the continuum electron operator in terms of lattice creation operators associated to an orbital $\alpha$ and a lattice site  $\mathbf{R}$, via $ c^\dag_{\mathbf{r}} = \sum_{\alpha\mathbf{R}} w_{\alpha\mathbf{R}}(\mathbf{r}) c^\dag_{\alpha\mathbf{R}}$ where $w_{\alpha\mathbf{R}}(\mathbf{r})$ are a set of real space orbitals indexed by $\alpha$ and centered at the lattice sites $\mathbf{R}$, e.g. \cite{kreisel2021quasi, sobral2023machine}. One obtains
\begin{align}
    \text{LDOS}(E,\mathbf{r}) &= - \tfrac{1}{\pi} \text{Im} \sum_{\alpha\alpha',\mathbf{R}\mathbf{R}'} w^*_{\alpha\mathbf{R}}(\mathbf{r}) w_{\alpha'\mathbf{R}'}(\mathbf{r}) \langle c_{\alpha\mathbf{R}} c^\dag_{\alpha'\mathbf{R}'}\rangle_E \nonumber\\
  &= - \tfrac{1}{\pi} \text{Im} \sum_{\alpha\alpha',\mathbf{R}\mathbf{R}'} w^*_{\alpha\mathbf{R}}(\mathbf{r}) w_{\alpha'\mathbf{R}'}(\mathbf{r}) G(E, \mathbf{R}-\mathbf{R}')_{\alpha\alpha'}
\end{align}
where $G(E, \mathbf{R}-\mathbf{R}')_{\alpha\alpha'}$ is the Green's function of the lattice model,
\begin{gather}
G(E, \mathbf{R}-\mathbf{R}')_{\alpha\alpha'} = \int \frac{d^d\mathbf{k}}{(2\pi)^d} \left(\frac{1}{E - \mathcal{H}(\mathbf{k}) + i0} \right)_{\alpha\alpha'} e^{i\mathbf{k}(\mathbf{R}-\mathbf{R}')}
\end{gather}
where $\mathcal{H}(\mathbf{k})_{\alpha\alpha'}$ is the tight-binding Hamiltonian, defined as a matrix acting on the basis of orbitals $\alpha$.

\subsection{Charge density in three-band model of WSe$_2$}
In this section, we present calculations of the local density of states (LDOS) in a widely-used minimal three-band model of WSe$_2$ \cite{Liu2013three-band}, comprised of $d_{z^2}$, $d_{xy}$, and $d_{x^2-y^2}$ orbitals located on the triangular lattice formed by the tungsten atoms. The model is given by
\begin{align}
\label{threebandH}
H^{\mathrm{NN}}(\boldsymbol{k})=\left[\begin{array}{ccc}
h_0 & h_1 & h_2 \\
h_1^* & h_{11} & h_{12} \\
h_2^* & h_{12}^* & h_{22}
\end{array}\right]
\end{align}
where
\begin{gather}
h_0=2 t_0(\cos 2 \alpha+2 \cos \alpha \cos \beta)+\epsilon_1, \\
h_1=-2 \sqrt{3} t_2 \sin \alpha \sin \beta+2 i t_1(\sin 2 \alpha+\sin \alpha \cos \beta), \\
h_2=2 t_2(\cos 2 \alpha-\cos \alpha \cos \beta)+2 \sqrt{3} i t_1 \cos \alpha \sin \beta, \\
h_{11}=2 t_{11} \cos 2 \alpha+\left(t_{11}+3 t_{22}\right) \cos \alpha \cos \beta+\epsilon_2, \\
h_{22}=2 t_{22} \cos 2 \alpha+\left(3 t_{11}+t_{22}\right) \cos \alpha \cos \beta+\epsilon_2 \\
h_{12}=\sqrt{3}\left(t_{22}-t_{11}\right) \sin \alpha \sin \beta +4 i t_{12} \sin \alpha(\cos \alpha-\cos \beta) \\
(\alpha, \beta)=(k_x a/2,  \sqrt{3}k_y a/2)
\end{gather}
with the hopping values for WSe$_2$ given by $t_0=-0.146$, $t_1=-0.124$, $t_2=0.507$, $t_{11}=0.117$, $t_{12}=0.127$, $t_{22}=0.015$, $\epsilon_1=0.728$, $\epsilon_2=1.655$ (all units in eV). We retained only nearest-neighbour hoppings; we shall present calculations with spin-orbit set to zero, to highlight that it is irrelevant to the essential physics of the $C_3$-enforced obstruction.

The tight-binding Hamiltonian accurately captures the mismatch of $C_3$ eigenvalues in the valence band -- the valence band wavefunction has a $C_3$ eigenvalue of $1$ near $\Gamma$ and $\omega = e^{2\pi i/3}$ near $K$. The difference in $C_3$ eigenvalues implies that the valence band is described by an obstructed atomic limit, meaning that it is not possible to construct a real-space basis for this band out of states exponentially localised on the atomic sites, using only the Bloch states from this band. Calculating the LDOS in this minimal model serves to illustrate that this key feature is the origin of the charge shifting, rather than any additional details captured in full DFT calculations of the charge density.

\begin{figure*}[t!]
    \centering
    \includegraphics[width=0.9\textwidth]{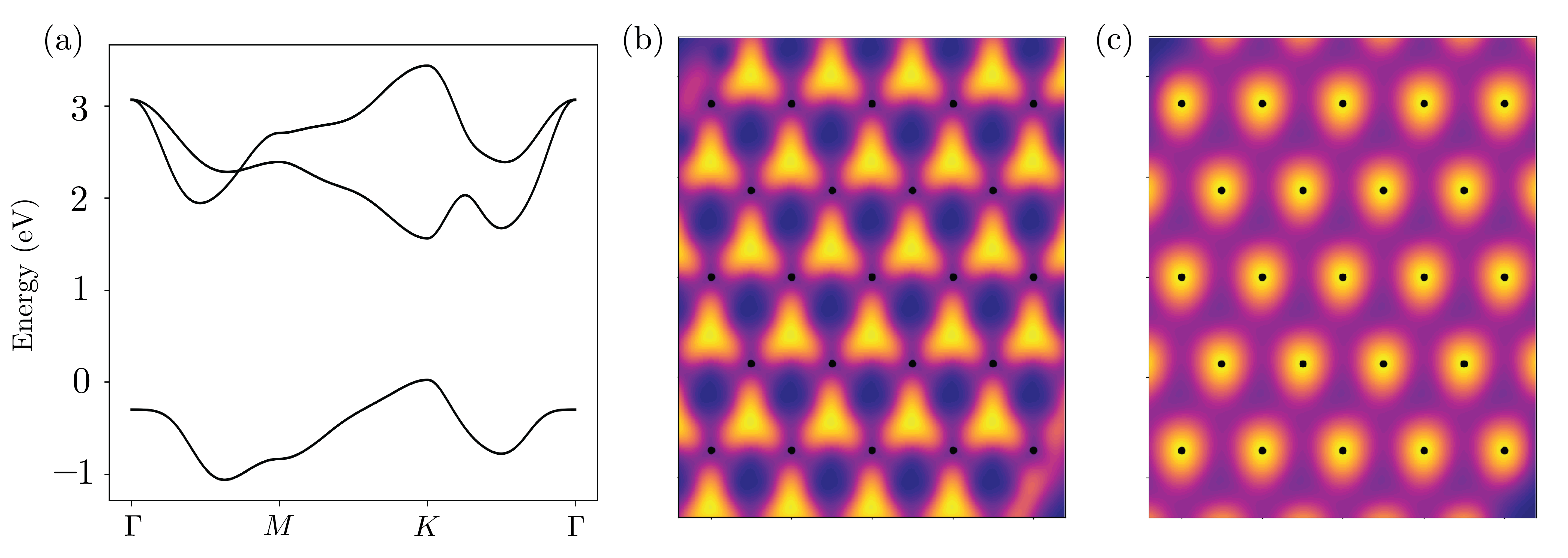}
    %[width=1\linewidth]
    \vspace{-0.1cm}
    \caption{ \textbf{Simulated local density of states in the three-band model of WSe$_2$.} (a) Band structure of WSe$_2$ in the minimal three-band model of Ref. \cite{Liu2013three-band}, Eq. \eqref{threebandH}. Note the neglect of spin-orbit coupling results in minor qualitative differences from Fig. \ref{f:fig3}c. (b) The local density of states in the three-band model near $E=0$, where the Fermi surface is close to the $K$-point, illustrating the localisation of charge in the center of the triangles (dark spots indicate the locations of the W atoms). (c) Local density of states near $E=-0.3$, where a sizable pocket near $\Gamma$ contributes, producing atomically-centered charge density.}
    \label{f:sm_fig1}
    \vspace{-0.1cm}
\end{figure*}

In Fig. \ref{f:sm_fig1}a, we show the resulting band structure with orbital weight overlaid. In this minimal model, at $0$ eV the chemical potential intersects only the $K$ pocket near the valence band maximum, while at $-0.3$ eV the chemical potential intersects the additional pocket at $\Gamma$. The density of states originating from the $K$ pocket is expected to live in the center of the triangular sites, while the contributions from $\Gamma$ live at the atomic sites and become more significant as the filling is lowered. The LDOS at these respective energies is shown in Fig. \ref{f:sm_fig1}b,c, in which one indeed observes a shift from the charge living in the centers of the triangular lattice to the atomic sites.

\begin{figure*}[t!]
    \centering
    \includegraphics[width=0.7\textwidth]{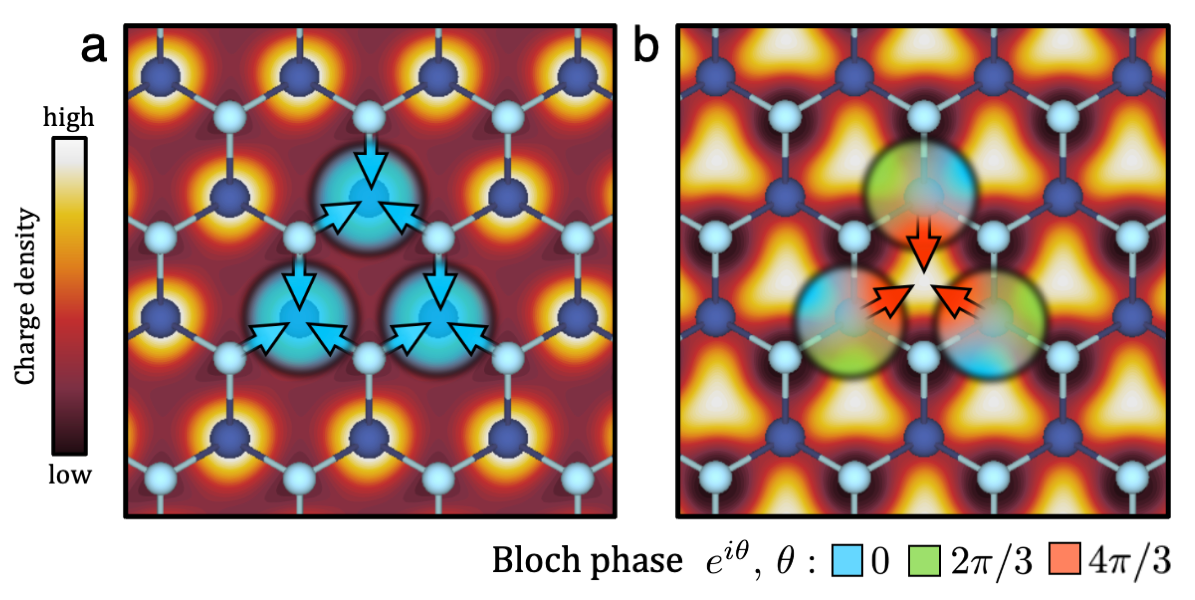}
    %[width=1\linewidth]
    \vspace{-0.2cm}
    \caption{ \textbf{Off-centered charge density from orbital interference.} In addition to the Wannier function viewpoint described in the main text Fig. \ref{f:fig2}, the off-centered charge density can be understood in terms of the overlap between the valence band and constituent atomic orbitals. (a) Near $\Gamma$, the orbital weight has trivial $C_3$ eigenvalue, and the charge density has the same phase everywhere near the atomic sites. (b) Near $K$, the valence band has $d+id$ character -- at each atomic site, the orbitals exhibit a phase winding indicated by the red/green/orange pinwheel; these phases add constructively near the hollow sites.}
    \label{f:sm_fig2}
    \vspace{-0.3cm}
\end{figure*}

In the main text, we explained this shift by describing the Bloch function as a linear combination of Wannier orbitals at the different sites. A complementary viewpoint is to consider the Bloch wavefunction as a linear combination of the original atomic $d$ orbitals, rather than the valence band Wannier function. Near $K$, the Bloch wavefunction possesses $d_{xy}\pm i d_{x^2-y^2}$ character; this chiral orbital undergoes a phase winding around the atomic sites: note that the in-plane angular dependence of this chiral combination of the $d$ orbitals takes the form $d_{xy}\pm i d_{x^2-y^2} \sim \cos2\theta \pm i \sin2\theta$. As shown in Fig. \ref{f:sm_fig2}, the phases add destructively at the atomic sites, but add constructively in the centers of the triangles. Hence, the off-centered charge density arises simply as a result of the phase winding of the Bloch function at the VBM (as has been discussed previously in the case of quantum spin Hall insulator indenene \cite{bauernfeind2021design, eck2022real}). It is interesting to note that the chirality of the VBM has a range of important ramifications for the optical properties of TMDs \cite{mak2016photonics}; our results highlight the manifestation of this orbital chirality in STM.

\subsection{Charge density in a two-band model of $s+f$ orbitals}

\begin{figure*}[t!]
    \centering
    \includegraphics[width=\textwidth]{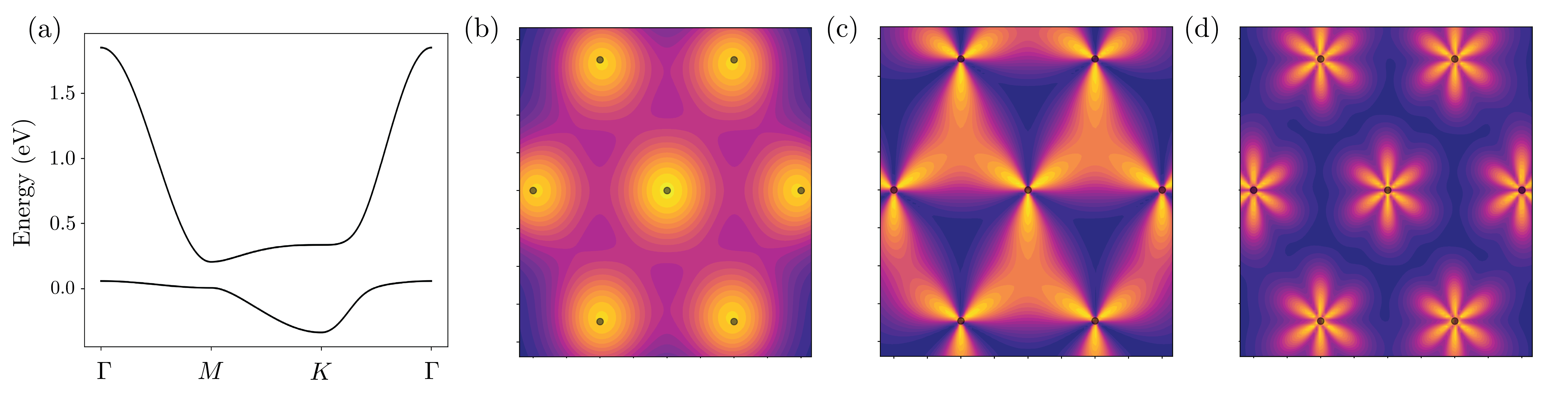}
    %[width=1\linewidth]
    \vspace{-0.5cm}
    \caption{ \textbf{Simulated local density of states in the $s+f$ orbital model.} (a) band structure of the $s+f$ orbital model defined in Eq. \eqref{sfH}. The model exhibits no $C_3$-enforced topological obstruction, as the transformation of the orbital weight under $C_3$ is the same at $\Gamma$ and $K$. (b) Near the $\Gamma$ point where the orbitals are predominantly $s$ character. (c) Near $K$, the band is an equally mixed $s+f$ orbital, which possesses a ``fidget spinner" like profile which strongly overlaps with neighbouring orbitals in the center of the triangles. One might therefore expect that the LDOS is concentrated in the center of the triangles. (c) LDOS of the $s+f$ model for $E=-0.33$ eV, where the Fermi surface is near $K$. The combination of the fidget spinner orbital with the Bloch phase $e^{i\mathbf{K}\cdot\mathbf{R}}$ results in destructive interference between neighbouring orbitals, canceling out the large contribution to the density of states at the triangle centers.}
    \label{f:sm_fig3}    
    \vspace{-0.1cm}
\end{figure*}

In this section we further clarify the origin of this effect using a fine-tuned two-band model to highlight that the spatial profile of the wavefunction at a given point in momentum space does not on its own determine the maximum of the charge density. A ``fidget spinner'' type charge profile can be produced by an equal combination of an $s$-orbital with an $f$-orbital $s+f$ shown in Fig. \ref{f:sm_fig3}c, possessing the angular dependence $\sim (1+\cos(3\theta))$. We construct a tight-binding model for which the valence band has pure $s$ character near the $\Gamma$ point, yet has $s+f$ character at the $K$ point.

The model is given by
\begin{align}
\label{sfH}
 \mathcal{H}(\mathbf{k}) = \begin{pmatrix}
t_s (3+\gamma(\mathbf{k})) & t_p\gamma(\mathbf{k}) \\
t_p\gamma(\mathbf{k}) &  t_f(3+\gamma(\mathbf{k}))\\
\end{pmatrix}
\end{align}
where $\gamma(\mathbf{k})=\cos(\mathbf{k}.\mathbf{a}_+)+\cos(\mathbf{k}.\mathbf{a}_-)+\cos(\mathbf{k}.(\mathbf{a}_+ +\mathbf{a}_-))$, and the $2\times 2$ matrix acts on the space of $(s,f)$ atomic orbitals; we shall choose $t_s = 0.00655$ eV, $t_f=0.2055$ eV, $t_p=-0.0646$ eV in what follows.

The band structure is shown in Fig. \ref{f:sm_fig3}a; the model is fine-tuned to produce pure $s$-orbital character near $\Gamma$ and an exactly equal mix $s+f$ orbital character near $K$. Fig. \ref{f:sm_fig3}c shows the real space profile of the $s+f$ orbital, showing maximal charge density at the centers of the triangles rather than on the atomic sites. Yet, the $C_3$ eigenvalue of the wavefunction at both $\Gamma$ and $K$ is trivial in this case, since both $s$ and $f$ orbitals are invariant under threefold rotations. Hence, the valence band in this model is not obstructed.

When we compute the local density of states at $\Gamma$, we find localised charge maxima on the atomic sites (Fig. \ref{f:sm_fig3}b), as expected. Yet, when we compute the LDOS near $K$, we do \textit{not} find a charge maximum living in the centers of the triangles. Rather, it is clear that the charge living in the center of the triangles has been eliminated by destructive interference, leaving only the portion of the $s+f$ orbital closest to the atomic site (Fig. \ref{f:sm_fig3}d). Without a non trivial $C_3$ eigenvalue from the orbital weight, the Bloch phase $e^{i\mathbf{K}\cdot \mathbf{r}}$ results in destructive interference at the triangle centers; it is therefore necessary for the orbital character to transforms non trivially under $C_3$ in order for the charge to be located in the centers of the triangles. 

This fact has already been discussed in theory work by Eck et al. \cite{eck2022real}, who observed that chiral $p_x\pm i p_y$ orbitals located on the $1a$ positions produce a charge maximum in the center of the triangles i.e. $2b$ positions, a result they use to explain the charge profile seen in STM imaging of indenene \cite{bauernfeind2021design}.

\section{Additional DFT calculations}
\begin{figure}[h!]
    \centering
    \includegraphics[width=0.6\linewidth]{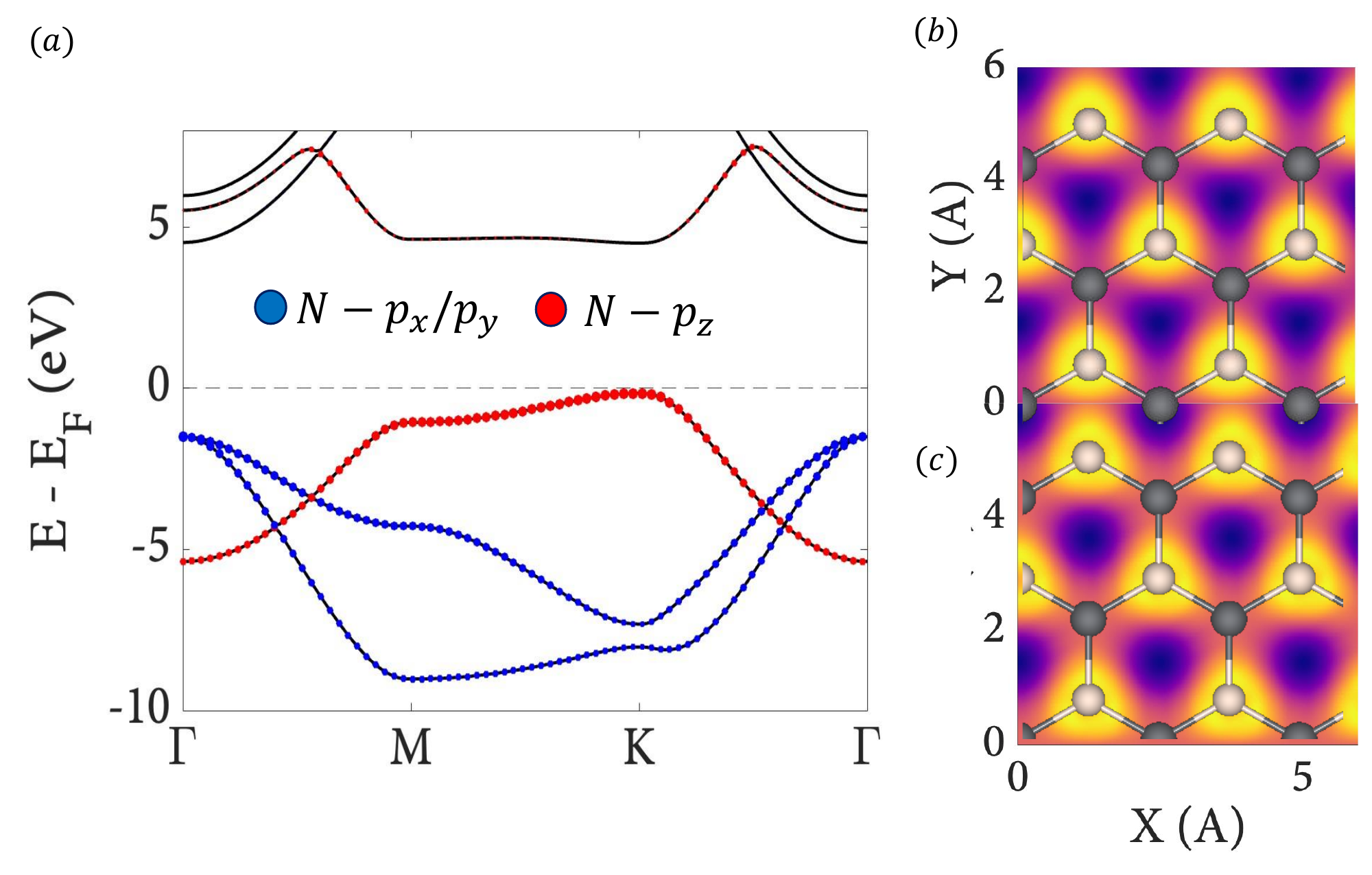}
    \vspace{-0.3cm}
    \caption{\textbf{Band structure and charge density of hBN for two bias voltages.} (a) Band structure with orbital projections. The valence band is mainly composed of $p_z$ orbitals, centered on the nitrogen sites. Both the $p_z$ and $p_x,p_y$ states which make up the valence manifold are extremely weakly hybridized, admitting maximally localized Wannier functions plotted in Fig. \ref{f:fig2} Colored circles represent the weight of atomic-like orbitals for $N$. (b) Charge density of states with $V_\textrm{bias} = -0.5V$, in the vicinity of the $K$ point. (c) Charge density at $V_\textrm{bias} = -1.5V$, covering bands near $\Gamma$. In both (b),(c), black and white atoms superimposed on the charge density denote boron and nitrogen, respectively. }
    \label{fig:hbn_bands}
\end{figure}

Here, we present the band structure and partial charge density of hBN as a function of bias voltage.  In Fig.~\ref{fig:hbn_bands}a we plot the band structure of hBN around the Fermi level. The valence band is primarily composed of $p_z$-like states, as shown in the orbital decomposition. We calculate the partial charge density up to a voltage $V_\textrm{bias}$. Unlike the case of the obstructed atomic limit in WSe\textsubscript{2} (as shown in the main text), the charge density in hBN does not evolve with bias voltage, as shown in Figs.~\ref{fig:hbn_bands}b- \ref{fig:hbn_bands}c. As we vary the voltage following the procedure carried out in the main text -- from states predominantly near $K$ to those near $\Gamma$ -- the charge density does not exhibit any movement in real space.  Orbital hybridization within the $p$ manifold of $N$ is much weaker than the corresponding $d$ manifold in WSe2\textsubscript{2}. 
The band composition of the band near $\Gamma$ includes $p_x$, $p_y$ orbitals, yet the states do not show any orbital interference due to the strongly localized nature of the wavefunction: the spread of the Wannier states $\sigma = 0.43 a_0$, where $a_0 = 2.51$ \AA  \ is the lattice constant of hBN. This demonstrates the strength of our approach, allowing for distinguishing between trivial (hBN) vs non-trivial states (WSe\textsubscript{2}).

\subsection{Height dependence of simulated images}

The shape of the charge density alone is insufficient to determine the topological properties. Below, we show that varying the tip height (the $z$ projection of the partial charge density in simulation) changes the shape of the imaged states, without revealing additional information about a possible obstruction.

The distinct shape of the simulated charge centers shown in the main text and in Fig.~\ref{fig:hbn_bands}(b)-(c) requires specification of the $z$ projection of the partial charge density, $\rho(x,y,z)$. An immediate question is whether the tip position, reflected in $z$ may evince any topological properties of the imaged charge.
\begin{figure}
    \centering
\includegraphics[width=0.8\linewidth]{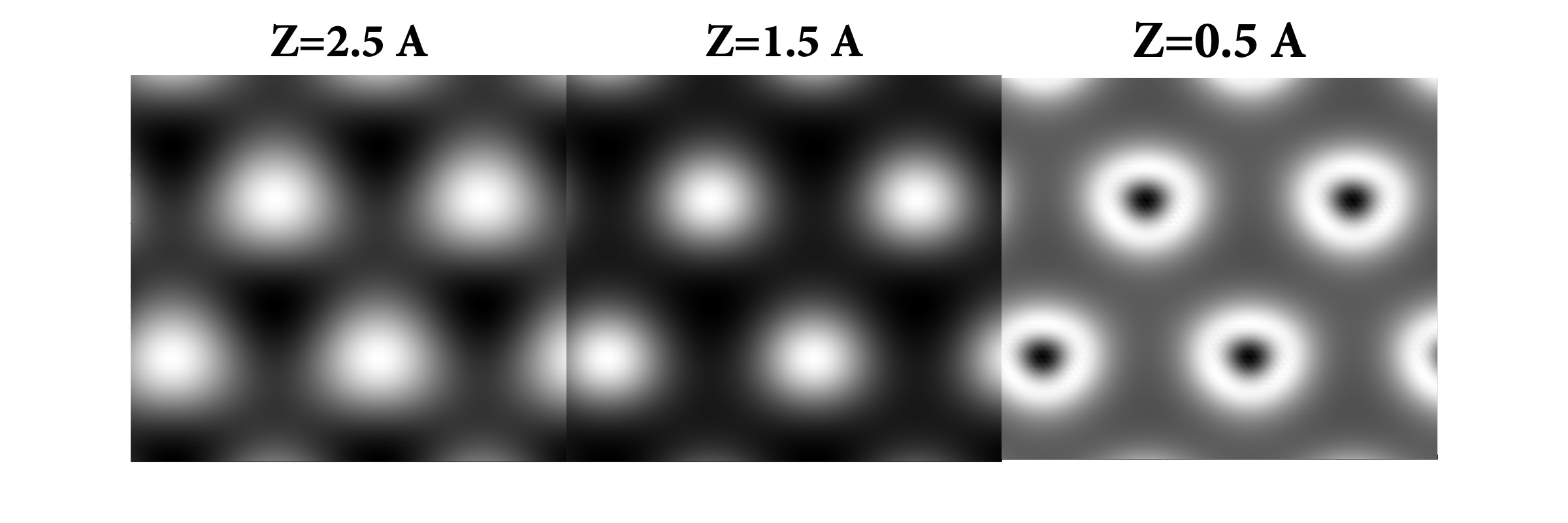}
\vspace{-0.3cm}
    \caption{\textbf{Varying the $z$ component of the partial charge density in WSe\textsubscript{2}.} The charge density in the vicinity of the $\Gamma$ point is shown for three different heights. The bias is set to the same value as that in the main text.}
    \label{fig:height}
\end{figure}
In Fig.~\ref{fig:height}, we vary the $z$ projection from $2.5\AA$ to $0.5\AA$ in imaging the charge density of WSe\textsubscript{2}. We find a gradual deformation of the orbital shape, until the annular pattern of the $d$-orbitals near $\Gamma$ is apparent. While the charge seems to evolve with $z$ height, it is clear that it is impossible to turn the topological nature of the states, even if an annular or off-atomic-site state appears when $z = 0.5\AA$. The precise determination of the topological properties requires probing the band at different energies, hence the unique advantage of our approach.

\section{Expanded Discussion on STM Measurements of WSe$_2$}

Here we include additional technical discussion regarding our strategy to measure the energy dependence of the local density of states.  One commonly used method is to perform $dI/dV$ spectroscopy maps, where a full spectrum is obtained at each pixel. This enables comparison of the LDOS at different energies at the same spatial position. However, this procedure is typically performed by establishing constant current feedback at each pixel, resulting in a variation of tip-sample height between pixels. The contrast in such spectroscopic maps can arise from this feedback artifact, and is not a reliable indicator of shifts in the probability density in our case. Another typical technique is to obtain constant current topography measurements at the same location with different bias conditions. In this case, it is difficult to compare the spatial features in images obtained consecutively, as lateral drift can occur between images, especially at room temperature. While both techniques are conceptually simple, these complicating factors must be accounted for. For this reason we determine that topography and current measurements where the bias is changed mid-frame is the most reliable way to capture the energy dependence of the LDOS.

\subsection{Supplemental description of the measurement procedure of the STM current image}

\begin{figure*}[t!]
    \centering
    \includegraphics[width=0.7\textwidth]{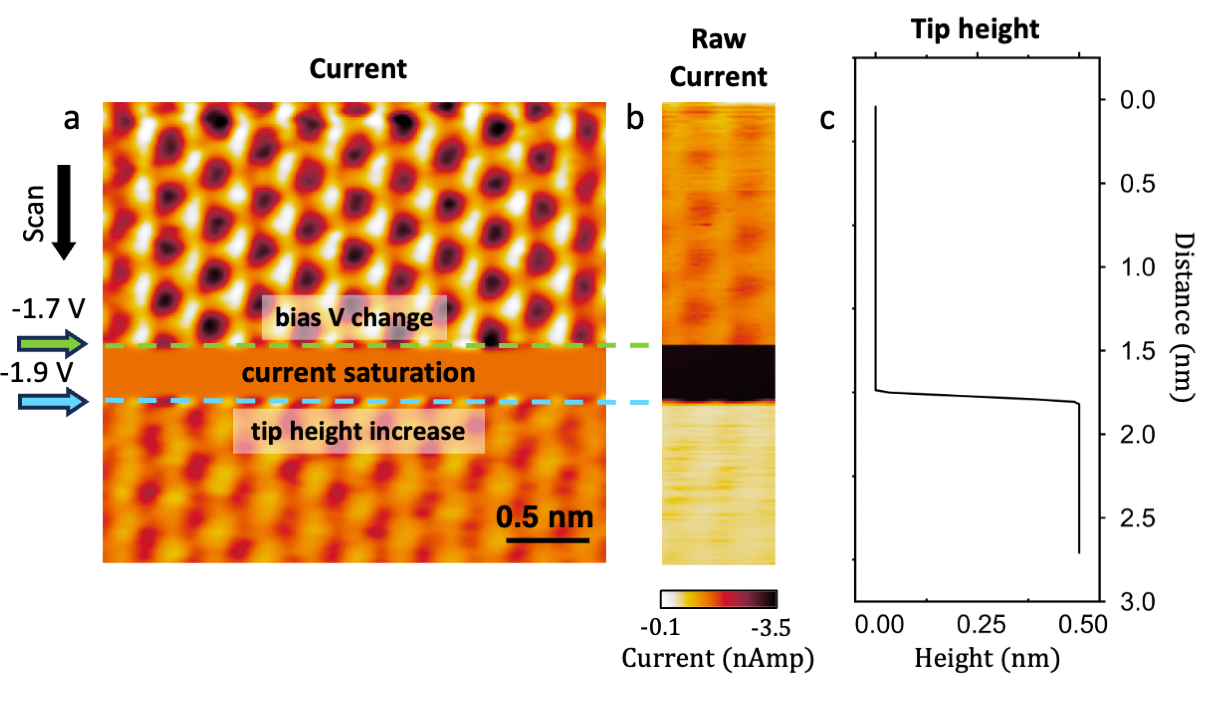}
    %[width=1\linewidth]
    \vspace{-0.1cm}
    \caption{ \textbf{Detailed description of STM current imaging procedure} (a) STM current image at constant tip height. The scan begins at -1.7 V (top), then the bias was changed to -1.9 V (green arrow and dashed line), and the tip height was increased by 0.5 nm (blue arrow and dashed line). The image was flattened to lower the contrast and better visualize the different regions. (b) The corresponding raw current image showing the quantitative changes in the tunneling current amplitude (c) A height profile showing the tip position at the different regions of the current scan.}
    \label{f:current}
    \vspace{-0.1cm}
\end{figure*}

In this section we outline the procedure for obtaining the current STM images included in Fig. \ref{f:fig3}f of the main text. Supplementary Fig. \ref{f:current}a shows an STM current image that begins at the top of the frame at a sample bias of -1.7 V with a constant tip-sample height (feedback disabled). As discussed in the main text, the tip must be closer to measure the tunneling current at this bias voltage because only states with a large tunneling decay constant near the K point  are probed. When the sample bias is changed to -1.9 V (green dotted line) the states near the gamma point are also probed, and the current is saturated due to a sharp increase in tunneling. Supplementary Fig. \ref{f:current}a is a flattened current image that clearly shows the different bias regions, and Supplementary Fig. \ref{f:current}b shows the corresponding raw current image that quantitatively shows the order of magnitude increase in the tunneling current amplitude. To resolve the lattice, the tip was moved away from the surface by 0.5 nm (blue dashed line) to decrease the current, as indicated by the corresponding height profile in Supplementary Fig. \ref{f:current}c. The same procedure was followed to obtain the current measurement in Fig. \ref{f:fig3}f of the main text, but the bias and tip height were adjusted simultaneously to produce a seamless image without current saturation.

\subsection{Defect analysis to determine the WSe\textsubscript{2} atomic lattice alignment}
\begin{figure*}[t!]
    \centering
    \includegraphics[width=0.6\textwidth]{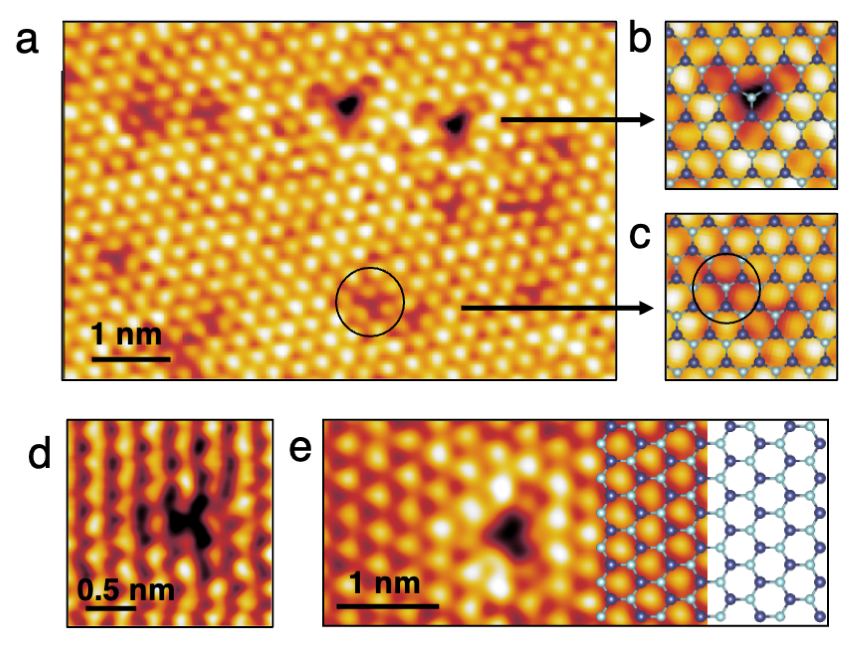}
    %[width=1\linewidth]
    \vspace{-0.2cm}
    \caption{ \textbf{Analysis of the WSe\textsubscript{2} lattice alignment with the STM images} (a) STM image of the S doped WSe\textsubscript{2}. (b) O substitutional impurity on an Se site. (c) S impurity on the Se site with a circle added to guide the eye. (d), (e) O impurities on the Se sites obtained near the STM images in Fig. \ref{f:fig3} of the main text. The WSe\textsubscript{2} lattice overlay shows the alignment determined from the Se site impurity positions. The sample bias in (a,b,c,e) is -1.8 V and (d) is -1.6 V.}
    \label{f:defect}
    \vspace{-0.1cm}
\end{figure*}

Here we describe the method we used to determine the alignment of the WSe\textsubscript{2} atomic lattice in the STM images in Fig. \ref{f:fig3} of the main text. Supplementary Fig. \ref{f:defect}a shows an STM image of the S doped WSe\textsubscript{2}, that shows the S dopants along with an additional defect type. These defects closely resemble the oxygen impurities substituting an Se site that were previously reported in MoSe\textsubscript{2} \cite{barja2019identifying}. Supplementary Figs. \ref{f:defect}b,c give a closer comparison of the two defect types. With the known lattice site of the S defects and knowing that STM probes the honeycomb center at a sample bias near the K point (as shown in Fig. \ref{f:fig1} of the main text), we can accurately align the atomic lattice in the images. This analysis shows that the unknown defects are located at the Se sites, supporting the conclusion that they are O impurities on the Se sites. We will refer to them as Se site impurities.
With this insight, we can use the Se site impurities to determine the lattice alignment of the STM images of the undoped WSe\textsubscript{2} monolayers in \ref{f:fig3}. Supporting Figs. \ref{f:defect}d,e show images of the Se site impurities obtained at locations near the STM images in \ref{f:fig3}. From these images we determine the location of the Se sites and align the WSe\textsubscript{2} atomic lattice. 

\end{document}